# Gate-tunable Rashba spin-orbit coupling and spin polarization at diluted oxide interfaces

Yulin Gan*, Yu Zhang, Dennis Valbjørn Christensen, Nini Pryds, and Yunzhong Chen*

*Department of Energy Conversion and Storage, Technical University of Denmark, Risø Campus, 4000 Roskilde, Denmark*

**Abstract:**

Diluted oxide interface of $LaAl_{1-x}Mn_xO/SrTiO_3$ (LAMO/STO) provides a new way of tuning the ground states of the interface between the two band insulators of LAO and STO from metallic/superconducting to highly insulating. Increasing the Mn doping level ($x$) leads to a delicate control of the carrier density as well as a raise in the electron mobility and spin polarization. Herein, we demonstrate a tunable Rashba spin-orbit coupling (SOC) and spin polarization of LAMO/STO ($0.2 \leq x \leq 0.3$) by applying a back gate. The presence of SOC causes the splitting of energy band into two branches by a spin splitting energy. The maximum spin splitting energy depends on the Mn doping and decreases with the increasing Mn content and then vanishes at $x = 0.3$. The carrier density dependence of the spin splitting energy for different compositions shows a dome-shaped behavior with a maximum at different normalized carrier density. These findings have not yet been observed in LAO/STO interfaces. A fully back-gate-tunable spin-polarized 2DEL is observed at the interface with $x = 0.3$ where only $d_{xy}$ orbits are populated ($5.3 \times 10^{12}$ cm$^{-2} \leq n_s \leq 1.0 \times 10^{13}$ cm$^{-2}$). The present results shed light on unexplored territory in SOC at STO-base oxide heterostructures and make LAMO/STO an intriguing platform for spin-related phenomena in $3d$-electron systems.

# I. INTRODUCTION

The two-dimensional electron liquid (2DEL) at the interface or surface of $SrTiO_3$ (STO) have attracted extensive attention due to the extraordinary physical properties such as spin-orbit coupling [1,2,3,4], ferromagnetism [5,6,7], superconductivity [8,9,10], and quantum Hall effect [11]. These properties can also be controlled through manipulation of the carrier concentration using electric fields or chemical doping. So far, the archetypical system remains the 2DEL formed at the interface between two band insulators of $LaAlO_3$ (LAO) and STO. Due to the large permittivity of the STO substrate at low temperature [12], the transport properties of the 2DELs such as carrier density and mobility can be modulated significantly by electrostatic gating [13,14]. Gate-tunable metal-insulator transitions [15], superconductivity [8], Rashba spin-orbit coupling [1] and spin polarization [16] have been reported, which could pave the way for developing future novel electronic/spin-electronic devices [17,18].

Based on the chemical doping and a manganite-buffer layer induced modulation doping [19,20], a new modulation doping made by alloying $LaMnO_3$ (LMO) with LAO not only suppresses the sheet carrier density ($n_s$) gradually by increasing the Mn doping level ($0 \leq x \leq 1$) but also leads to the realization of spin-polarized 2DELs at $x = 0.3$. [21] This magnetically diluted oxide interface reveals the evolution of electronic state at STO-based interfaces as a function of carrier density and provides new and unexplored electronic states than the intensively investigated bare LAO/STO interface [8,22]. Four kinds of electronic ground states have been created at diluted interfaces by controlling the Mn doping level. i) For $0 \leq x \leq 0.2$, conductivity occurs in two bands; ii) for $0.225 \leq x \leq 0.275$, only a single band is occupied; iii) for $x = 0.3$, the single band occupancy remains, however, the 2DEL becomes spin-polarization; iv) for $0.3 < x \leq 1$, an insulating state is observed. This provides a convenient handle to investigate the gate-tunable transport properties of different as-grown ground states.

Generally, the 2DEL of STO-based interfaces is confined in an asymmetric quantum well located on the STO side. The absence of spatial inversion symmetry at the heterointerfaces leads to a strong Rashba spin-orbit coupling (SOC) [1]. In 2D electron systems, weak localization (WL) and weak anti-localization (WAL) are two quantum interference effects resulting from electron phase coherence and spin-orbit coupling. These quantum interference effects are related to the three scattering processes - inelastic scattering, spin orbit scattering and elastic scattering [23]. Previous studies have shown that the SOC strength can be modulated by top- and/or back-gate [1,3,4,22,24]. Analyzing the WAL/WL, some groups reported that the strength of the SOC decreases monotonously upon decreasing the back-gate voltages [1,10] or even show a maximum at the $d_{xy}$-$d_{xz}/d_{yz}$ crossing region at LAO/STO interface [3]. Previous studies of the SOC at STO-based conducting interfaces implemented the weak SOC approximation [25], which assumes that the elastic scattering characteristic field is much stronger than the SOC scattering characteristic field. A significant mobility ($\mu$) enhancement at the LAMO/STO interfaces is observed when increasing the Mn doping and leads to a decrease of the elastic scattering field. When the elastic scattering field is so weak that it is comparable with the SOC scattering field, the weak SOC approximation is not anymore applicable and it remains to be investigated what happens to the spin orbit effect here. Our LAMO/STO system exhibits different ground states with the possibility to enhance the mobility by increasing the Mn doping level. The LAMO/STO system, therefore, provides an interesting platform to investigate the gate-tunable SOC where the weak SOC approximation is not applicable.

Moreover, the anomalous Hall effect (AHE) observed in the LAMO/STO interface with $x = 0.3$ implies the existence of a spin-polarized state with a single band of $d_{xy}$ electrons. In previous report, D. Stornaiuolo *et al.* [16] has shown an electric-field-tunable spin-polarized and superconducting 2DELs at the ferromagnetic EuTiO$_3$ buffered LAO/STO interface with higher carrier density where both the $d_{xy}$ and $d_{xz}/d_{yz}$ bands are populated. The AHE observed in the current work is a result of spin-polarized $d_{xy}$ electrons, making our system interesting to investigate the gate-controlled magnetism or spin polarization.

In this paper, we systematically investigated the back-gate-tunable transport properties of various electronic ground states at LAMO/STO heterostructures for different Mn doping ($0.2 \leq x \leq 0.3$). We observed a gate-tunable SOC in the heterointerfaces with $x = 0.2$, $0.225$ and $0.25$ and gate-tunable spin-polarized $d_{xy}$ electrons for $x = 0.3$. For $x = 0.2$, the weak SOC approximation is available and 2DEL exhibits a dome-shaped dependence of spin splitting energy on carrier density with the maximum occurring at the Lifshitz transition point. However, this approximation cannot be applied to analyze the WAL for the $x = 0.225$ and $0.25$, since the elastic scattering is comparable with the SOC scattering. In addition, the strongest spin splitting energy of different samples decreases with the increasing Mn doping level. On the other side, the spin-polarization of $d_{xy}$ electrons is fully tunable by the electric field effect at $x = 0.3$ with $5.3\times10^{12}$ cm$^{-2}$ $\leq n_s \leq 1.0\times10^{13}$ cm$^{-2}$.

## II. EXPERIMENTAL METHOD

The LaAl$_{1-x}$Mn$_x$O$_3$ thin films, with $x = 0.2$, $0.225$, $0.25$ and $0.3$, were grown by pulsed laser deposition (PLD) using a KrF laser on TiO$_2$-terminated $5\times5\times0.5$ mm (001) STO substrates. Before the deposition, a hard mask made from amorphous LaMnO$_3$ (LMO) was patterned on the STO substrates in a Hall bar geometry by optical lithography ($W = 50$ μm, $L = 500$ μm) [26]. The film thickness was kept constant at 8 unit cells, and the film growth of the unpatterned sample was realized in a layer-by-layer 2D growth mode as described in Ref. [21]. A pulsed laser with a wavelength of 248 nm, a repetition frequency of 1 Hz, and laser fluence of 4.0 J/cm$^2$ was used to ablate the LAMO ceramic targets. Films were deposited at 650 ℃, in $1\sim3\times10^{-5}$ mbar of O$_2$ and a constant distance between target and substrate of $\approx 50$ mm. After the growth of the film, the samples were cooled under the deposition pressure at a rate of 15 ℃/min to room temperature.

The electrical contacts to the interface of the Hall Bar samples were made using ultrasonically wire-bonded aluminum wires as electrodes. A uniform thin layer of silver paint is applied on the back of the substrate as the back-gate electrode. A CRYOGENIC cryogen-free measurement system was employed to characterize the magnetotransport at 2 K with various back-gate voltages ($V_{bg}$) and perpendicular magnetic field up to 15 T. The $V_{bg}$ is initially ramped to the highest positive value (80 V for $x = 0.2$, 0.225 and 0.3; 40 V for $x = 0.25$) and is then swept in the negative direction with steps of 10 V while measuring the $V_{bg}$-dependent transport characteristics until the metal-insulator transitions occur. The $V_{bg}$-dependent sheet resistances of all samples are shown in Figure S1 (Supporting Information). At all cases, the leakage current was less than 1 nA and the LMO hard masks are totally insulating.

## III. RESULTS AND DISCUSSION

Figure 1(a) displays a sketch of back-gated LAMO/STO interface patterned in the Hall-bar geometry. A back-gate bias, $V_{bg}$, is applied across the insulating STO substrate. Based on the phase diagram obtained in Ref. [21], we have chosen four representative compositions of $x = 0.2$, 0.225, 0.25 and 0.3, in order to investigate the effect of back-gate field on the interfacial conductivity of various ground states. The corresponding as-grown properties at 2 K are listed in Table 1, and the temperature-dependent sheet resistances are shown in Figure 1(b). All interfaces show typical metallic behavior. For $x = 0.2$, the electrons are populated in both the $d_{xy}$ and the $d_{xz}/d_{yz}$ bands, showing a typical two-band behavior. When $0.225 \leq x \leq 0.3$, only $d_{xy}$ electrons contribute to the interfacial conduction because the electrons in the $d_{xz}/d_{yz}$ bands are depleted. Moreover, at $x = 0.3$, the interfacial 2DEL becomes spin-polarized due to the proximity effect from ferromagnetic LAMO layer.

The Hall resistance ($R_{xy}$) measurements at different back-gate voltages reveal an effective modulation of the carrier density. The back-gate controlled Hall effect can be divided into the three regimes described below: i) For $0.2 \leq x \leq 0.225$ under all $V_{bg}$ investigated and $x = 0.25$ at $-10$ V $\leq V_{bg} \leq 10$ V, the Hall effect is suggestive of ordinary Hall effect (OHE) with one or two kinds of carriers, as shown in Figure 2(a-c); ii) For $x = 0.25$ at $20 \leq V_{bg} \leq 40$ V, the Hall resistance curves show the feature of two-carrier transport with AHE, which are discussed in detail in the supporting information (Figure S6); (iii) For $x = 0.3$ shown in Figure 3a, the Hall resistance shows a typical behavior of AHE with only one type of carrier. The Hall effect is linear at lower back-gate biases, whereas above a critical voltage ($V_c$) the Hall effect becomes nonlinear with an anti-clockwise bending of the curve at high field. For $0.2 \leq x \leq 0.225$ under all $V_{bg}$ investigated and $x = 0.25$ at $-10$ V $\leq V_{bg} \leq 10$ V, the linear and nonlinear Hall curves originate from the OHE with one-carrier and two-carrier transports, respectively, analogous to Hall effect reported in previous works [9,21]. Such transition from a linear to a nonlinear behavior is interpreted as an evidence of a Lifshitz transition as the carrier occupation changes from only the $d_{xy}$ bands to both the $d_{xy}$ and $d_{xz}/d_{yz}$ bands. Previous theoretic [27] and experimental reports [9] suggest that the tetragonal distortion and quantum confinement along $z$ direction push the in-plane $d_{xy}$ bands down in energy compared to the out-of-plane $d_{xz/yz}$ bands. Therefore, when the Fermi surface enters the bottom of the $d_{xz/yz}$ bands, both the $d_{xy}$ and $d_{xz/yz}$ bands become populated. Following the two-band conduction model in Ref. [9,28], we extract the electron densities $n_{LF}$ and $n_{HF}$ from the Hall coefficients ($R_h = dR_{xy}/dB$) near 0 T and 15 T, which reflect approximately the carrier density in the high mobility band $n_{LF} = 1/(eR_h(B \to 0))$ and the total carrier density $n_{HF} = 1/(eR_h(B \to \infty))$, respectively. Because the presence of AHE in $x = 0.25$ at $20 \leq V_{bg} \leq 40$ V leads to a correction to Hall resistance at low field, the $n_{LF}$ are extracted after removing the correction from AHE (Figure S6, Supporting Information).

A gate-induced Lifshitz transition is clearly demonstrated in the $V_{bg}$-dependent carrier density as shown in Figures 2(d-f). The $n_{LF}$ and $n_{HF}$ are approximately identical at lower back-gate voltages, which is in agreement with a linear Hall resistance where only the $d_{xy}$ orbits are populated. Above the critical voltage ($V_c$), $n_{LF}$ shows a clear deviation from $n_{HF}$, which is the typical character of nonlinear Hall resistance expected from electron transport in both $d_{xy}$ and $d_{xz}/d_{yz}$ bands. The critical values of $V_c$ for $x$ = 0.2, 0.225 and 0.25 are 0 V, 70 V and 10 V, respectively. The corresponding Lifshitz density ($n_c$) is $2.3 \times 10^{13}$ cm$^{-2}$, $2.7 \times 10^{13}$ cm$^{-2}$ and $1.9 \times 10^{13}$ cm$^{-2}$, respectively, which are similar to the reported value of $n_c \approx 1.7 \sim 3.1 \times 10^{13}$ cm$^{-2}$. [3,9,21] The minor differences in the critical carrier density among different samples might be due to the gate-dependent effective band structure. [22]

Furthermore, for the interface with $x$ = 0.3 hosting the spin-polarized 2DEL, the Hall resistance traces at different $V_{bg}$ shown in Figure 3(a) are nonlinear with a clockwise bend at low fields, which are different from those of the two-band conduction but similar to AHE with one-carrier transport reported in Ref. [21]. Here, the evolutions of Hall coefficients $R_h$ as a function of $B$ in Figure 3(b) further suggest that the $B$-dependent Hall resistances are nonlinear at low magnetic field and become linear at higher magnetic field, forming a peak around $B$ = 0. Such nonlinear behavior of $R_{xy}$ cannot be modeled by the typical two-band model, but rather by AHE with only $d_{xy}$ electrons. Therefore, such nonlinear Hall resistance is a sum of two contributions: the linear contribution from the OHE with one-carrier transport and a non-linear contribution from the AHE. The anomalous Hall resistance ($R_{xy}^{AHE}$) can be described using the Langevin-type function. [29] The total Hall resistance can then be expressed as $R_{xy} = R_{xy}^{OHE} + R_{xy}^{AHE} = -B/(en) + R_{sat}^{AHE} \tanh(B/B_c)$. Here, the $B_c$ is the critical field above which $R_{xy}^{AHE}$ saturates to the value of $R_{sat}^{AHE}$. The $R_{sat}^{AHE}$ is proportional to the saturation magnetization.

The fit of the AHE curve successfully extracts the ordinary Hall resistance ($R_{xy}^{OHE}$) and anomalous Hall resistance ($R_{xy}^{AHE}$). From the ordinary Hall resistance, the carrier density for different $V_{bg}$ is extracted and is used to calculate the mobility. As shown in Figure 3(c), with the increase of the $V_{bg}$, the carrier density is changed from $5.3 \times 10^{12}$ cm$^{-2}$ at 40 V to $1.0 \times 10^{13}$

cm$^{-2}$ at 80 V, and the mobility increases by almost six times from 420 cm$^2$/Vs to 2503 cm$^2$/Vs. It is worth noting that the carrier density of all samples could not return to their as-grown states after removing the gate voltage, which is common behavior in oxide interface systems [8,24]. This leads to a sharp $V_{bg}$-induced metal-insulator transition under 40 V at $x = 0.3$. The as-grown data are shown in Figure 2(d-f) and Figure 3(c) as open symbols, whereas the results after cycling the back-gate are shown as filled symbols. A possible reason for this behavior is that with the application of positive gate bias, some electrons are trapped in defect at the in-gap states [30]. On the other side, the $B$-dependent anomalous Hall resistances at various $V_{bg}$ shown in Figure 3(d) are extracted by deducting the ordinary Hall resistance from total Hall resistance. As summarized in Figure 3(e), the value of $R_{sat}^{AHE}$ increases from 24.0 Ω to 58.1 Ω as $V_{bg}$ decreases from 80 V to 50 V, and finally it decreases to 36.0 Ω at $V_{bg} = 40$ V. Our results suggest that the spin polarization of $d_{xy}$ electrons in our system is fully gate-tunable. And the decrease of the $R_{sat}^{AHE}$ at $V_{bg} = 40$ V might indicate the presence of the Kondo effect [31]

We now turn to the gate modulation of the longitudinal resistance. The magnetoresistance (MR: $\Delta R_{xx} = R_{xx}(B)/R_{xx}(0) - 1$) of $x = 0.2$, 0.225 and 0.25 under different electric fields are shown in Figures 4(a-c) as a function of $B$. At back-gate voltages higher than 50 V for $x = 0.2$ and 70 V for $x = 0.225$, the MR is positive and quadratic at all measured magnetic fields, which is a typical behavior of ordinary MR primarily caused by the classical Lorentz force. When $V_{bg}$ is below these values, a sharp increase of MR starts to appear at the low field and then drops after reaching a peak with increasing field. This phenomenon is a typical behavior of WAL, where the SOC counteracts the WL to form a peak superimposed on the negative magnetoresistance background caused by WL. For $x = 0.25$ at -10 V, only a negative MR is detected up to the largest accessible magnetic field (15 T), which is the typical signature of WL, resulting in a $V_{bg}$-induced WAL-WL transition. Therefore, every MR curve can be considered as a superposition of three components in different proportions, i.e. a sharp increase of MR around the zero field, negative MR and the quadratic and positive MR, which come from the WAL, WL and classical MR from Lorentz force, respectively.

The WAL/WL is a manifestation of the quantum interference behaviors of electrons, which is normally determined by inelastic (phase breaking) scattering, elastic scattering, and SOC scattering. In order to further analyze the quantum correction of WAL/WL to the MR, three characteristic magnetic fields ($B_k = \hbar/(4eD\tau_k)$, $k = i, e, so$) are introduced to characterize $B$-dependent quantum correction based on three scattering processes. Here, $\tau_i$, $\tau_e$ and $\tau_{so}$ are the inelastic scattering time, the elastic scattering time and the spin-orbit scattering time, respectively. The $D = v_F^2 \tau_e/2$ is the diffusion constant given by the Drude model. The elastic scattering time $\tau_e$ describes the characteristic time of an electron in an available closed path for electron backscattering. The inelastic scattering time $\tau_i$ is the characteristic time for the electron to avoid inelastic phase breaking. [32] When the elastic scattering time ($\tau_e$) of electrons is much less than the inelastic scattering time ($\tau_i$), i.e. $\tau_e \ll \tau_i$ and $B_e \gg B_i$, in absence of a significant SOC, the interference of backscattered electron waves is constructive, leading to the weak localization, such as observed at the interface with $x = 0.25$ at -10 V in Figure 4c. In the WL, the electrons are at localized states. Conversely, if $\tau_e > \tau_i$ (i.e. $B_e < B_i$), the elastic scattering time is so long that the electrons no longer return to their origin to cause a constructive interference, where the electrons are at extended states. For the extended states, the MR is mainly caused by the Lorentz force, such as observed in the interfaces with $x = 0.2$ at 50 ~ 80 V and $x = 0.225$ at 70 ~ 80 V as shown in Figures 4(a,b). This means that the WL takes place in fields $B_e \gg B_i$ (i.e. $\tau_e \ll \tau_i$), namely in the diffusive regime. On the other hand, the spin-orbit scattering time $\tau_{so}$ describes the spin phase shift on the order of $\pi$ caused by the effective magnetic field of the SOC [32]. When the SOC appears in the diffusive regime with $B_e > B_{so} > B_i$, the spin of carriers would add an extra phase factor that destroys the constructive interference, leading to the destructive interference to counteracts the WL - this is the character of WAL. Here, the $B_e$ could be rewritten as $B_e = e/(2hn_s\mu^2)$, which is only contain the carrier density and the mobility. As shown in Figure 4(d), the effective mobility ($\mu_{tot} = 1/[en_{tot}R_{xx}(0)]$) decrease when the carrier density ($n_{tot}$) is reduced, which is the same with the case of $x = 0.3$ shown in Figure 3(c). Notably, the samples with higher Mn always show higher mobility for the same carrier density. The deduced $B_e$ are shown in Figure 4(e) as a function of sheet carrier density normalized by the Lifshitz density, ($n_s/n_c$). Here, the adoption of normalized carrier density ($n_s/n_c$) effectively distinguishes the contributions of one-band region or two-band region. The $0 < n_s/n_c < 1$, $n_s/n_c = 1$ and $n_s/n_c > 1$ represent the one-band region, Lifshitz transition point and two-band region, respectively. For all samples, the back-gate controlled $B_e$ decreases monotonically with increasing normalized carrier density, due to the monotonically increasing mobility and sheet carrier density.

In the magnetoresistance measurements of the WAL/WL, the applied magnetic field ($B$) could break the time-reversal symmetry and diminish the interference of the time-reversed paths, leading to the compensation of the SOC and WL. As shown in Figures 4(a-c), for lower $B$ ($B_i < B < B_{so} < B_e$) the MR raises sharply with an increase of $B$, where the SOC reduces the net interference contribution from WL to form WAL. With the further increase of $B$, the MR reaches a maximum at $B_p$ after which the spin-orbit coupling cannot contribute to the quantum interference, marking the $B$-induced WAL-WL transition [33]. With a further increase of $B$, when $B_i < B_{so} < B < B_e$, the MR continues to decrease, due to the suppression of WL. Finally, when the $B$ is so high, i.e. $B_i < B_{so} < B_e < B$, quantum interferences are negligible, and only the classical positive MR from Lorentz force is detectable. Therefore, the magnetic field $B_p$ signifies the critical field for the $B$-induced WAL-WL transition, which roughly reflects the SOC strength. [3,34] The critical fields ($B_p$) are extracted directly from the clear peaks in WAL curves. However, for $x = 0.2$ at 0 V ~ 40 V, there is no observed peak because the weak WAL is overlapped with the classic MR. The corresponding $B_p$ is extracted after removing the classic MR fitted by the two-band model in Ref. [21]. (Figure S2, Supporting Information) As shown in Figure 4(f), the values of $B_p$ changes between 0.5 ~ 4.3 T, 1.2 ~ 2.1 T and 0.2 ~ 1.3 T for $x = 0.2$, 0.225 and 0.25, respectively, which are comparable to the observations in earlier work where the critical fields were found to be 0.3 ~ 4.0 T [3]. For $x = 0.2$, with the accumulation of the carriers, the $B_p$ first reaches a maximum value at the Lifshitz point and then decreases. This is similar to the nonmonotonically tunable spin-orbit coupling in STO-based heterointerfaces reported in Ref. [3]. Here, this strongest SOC observed at the Lifshitz transition is interpreted as an enhanced SOC due to the orbit hybridization at the $d_{xy}$-$d_{xz}$/$d_{yz}$ crossing region. However, for $x = 0.225$ and 0.25, as the carriers are reduced, the values of $B_p$ increase to a maximum and then decreases sharply or disappears just before a sharp metal-insulator transition occur. Interestingly, the maximum values of $B_p$ appear at $n_s/n_c = 0.52$ for $x = 0.225$ and at $n_s/n_c = 0.76$ for $x = 0.25$, where the Fermi-level is located in the deeper $d_{xy}$ band. Besides, no enhanced SOC occurs at the Lifshitz transition points. These phenomena at $x = 0.225$ and 0.25 are different from the observation at $x = 0.2$. The observed maximum $B_p$ for different doping levels decrease from 4.3 T to 1.3 T with the increase of Mn content from 0.2 to 0.25.

The shape of $B_p$ presented above reveals qualitatively the evolutions of SOC as a function of carrier density. As the consequence of SOC, the electron energy state at the Fermi surface splits into two branches characterized by a spin splitting energy. In order to analyze the characteristic field of SOC and the spin splitting energy in more details, the fit of the magnetoconductance (MC) is carried out to estimate the contributions from the WL, SOC and classical MC from Lorentz force. The MC ($\Delta\sigma_{xx} = \sigma_{xx}(B) - \sigma_{xx}(0)$) normalized by a universal value of conductance ($G_0 = e^2/\pi h \approx 1.2 \times 10^{-5}$ S) are shown in Figures S4(a-c) as a function of $B$ for various $V_{bg}$. Because of the larger quadratic field dependence of MC, for weak SOC in MR, the WAL in $\sigma_{xx}(B)$ is too weak to extract the SOC in samples with $x = 0.2$ at $30 \text{ V} \leq V_{bg} \leq 40 \text{ V}$ and $x = 0.25$ at $20 \text{ V} \leq V_{bg} \leq 40 \text{ V}$ (Figure S3, Supporting Information). Therefore, the significant SOC mainly emerges for $x = 0.2$ at $-40 \text{ V} \leq V_{bg} \leq 20 \text{ V}$ ($0.8 \leq n_s/n_c \leq 1.6$), for $x = $

0.225 at 10 V $\leq V_{bg} \leq$ 60 V (0.5 $\leq n_s/n_c \leq$ 0.9) and for $x$ = 0.25 at 0 V $\leq V_{bg} \leq$ 10 V (0.8 $\leq n_s/n_c \leq$ 1.0). Previous studies [1,2] show that the WAL at conductive STO-based interfaces is well described by two theoretical models developed by Hikami, Larkin, and Nagaoka (HLN) [35,36] and Iordanskii, Lyanda-Geller, and Pikus (ILP), respectively [37]. However, these models are valid only for a weak SOC in the diffusive regime, where the WL/WAL takes place at $B << B_e$ and $B_{so} << B_e$ for which the $B_e$ does not appear explicitly [38]. These methods give good fits for all WAL/WL curves, but the shape of the fitting parameter $B_{so}$ is in accordance with the prediction from $B_p$ only for $x$ = 0.2 (More detail discussions in Supporting Information, Section 3.). Because the extracted $B_{so}$ exceeds $B_e$ for $x$ = 0.225 at 40 V $\leq V_{bg} \leq$ 60 V and for $x$ = 0.25 at 10 V as seen in Figure S4(e,f), the two methods are not available for $x$ = 0.225 and 0.25, which indicates the failure of the weak SOC approximation. In this scenario, the strong SOC observed at the interfaces with higher mobility leads to the fact that $B_e$ is comparable with the $B_{so}$, so the weak SOC approximation fails to describe the SOC in WAL at $x$ = 0.225 and 0.25.

Here, the Maekawa-Fukuyama (MF) theory [25] is applied for the case of strong SOC to analyze the WAL/WL. Considering a pronounced contribution from Lorentz force to MC in the entire magnetic field regime, the classical MC gives rise to a parabolic behavior [10,39]. Combining the first order quantum correction of WL/WAL and the classical MC, the correction to the MC with a negligible Zeeman effect can be expressed as:

$$\frac{\Delta \sigma_{xx}(B)}{G_0} = -F(\frac{B}{B_1}) + F(\frac{B}{B_2}) + \frac{1}{2}F(\frac{B}{B_3}) - \frac{1}{2}F(\frac{B}{B_4}) - A_k \frac{\sigma_{xx}(0)}{G_0} \frac{B^2}{1+CB^2}$$

$$B_1 = B_e + B_{so}$$

$$B_2 = B_i + (B_e + B_{so})\frac{B_{so}}{B_e}$$

$$B_3 = B_i + (B_e + B_{so})(\frac{2B_{so}}{B_e - B_{so}})$$

$B_4 = B_i$

Here, the function $F$ is defined as $F(x) = \ln(x) + \psi(1/2+1/x)$, where $\psi(x)$ is the digamma function. The parameters $A_k$ and $C$ from the last term describe the classical MC. Using the $B_e$ deduced from the carrier density and mobility, the $B_{so}$ and $B_i$ are obtained directly from the best fits of the experimental data (Figure S4, Supporting Information). The shapes of $B_{so}$ for all samples are in accordance with the prediction from $B_p$ in Figure 4(f), indicating that the MF model successfully capture our observation of the WAL/WL. The SOC characteristic fields are tuned between 0.19~1.14 T, 0.09~0.31 T and 0.10~0.16 T for $x$ = 0.2, 0.225 and 0.25, respectively. Since the SOC contribution mainly stems from $d_{xy}$ electrons, we can derive the relaxation time $\tau_e$, $\tau_{so}$ and $\tau_i$ by assuming an average effective mass $m^* = m_e$ [11,40]. According to the expression in Ref. [1], the spin splitting energy $\Delta = \hbar/\sqrt{\tau_e \tau_{so}}$ is determined by the elastic scattering time and SOC scattering time, and its evolutions are summarized in Figure 5(a) as a function of normalized carrier density. The interface with $x$ = 0.2 has a domelike dependence of $\Delta$ on the carrier density, with the peak of 5.5 meV at the Lifshitz point ($n_s/n_c$ = 1). The significant local orbital angular momentum induced by the orbital hybridization at the $d_{xy}$-$d_{xz}/d_{yz}$ crossing area could enhance the SOC and form a dome. [3,41] Here, for one-carrier region ($0 < n_s/n_c \leq 1$), the $\Delta$ decreases monotonically with decreasing carrier density, similar to what observed in previous works [3,10]. In contrast, the spin splitting energies of $x$ = 0.225 and 0.25 increase monotonically with the decreasing carrier density, reaching the maximum values of 2.2 meV for $x$ = 0.225 at $n_s/n_c$ = 0.52 and 1.6 meV for $x$ = 0.25 at $n_s/n_c$ = 0.76, respectively. This was not discussed for the STO-based oxide interfaces so far. A possible explanation for our observations is that for $x$ = 0.225 and 0.25, the higher mobility causes a lower $B_e$, which is comparable with the $B_{so}$, so the weak SOC approximation ($B_{so} \ll B_e$) will fail to describe such SOC in WAL. At the same time, with the increasing carrier density, the decreasing $B_e$ reduces the probability of electron backscattering.

Because the WAL originates from the SOC of the backscattered electrons, the reduction of backscattered electrons further causes the suppression of spin-orbit coupled electrons, when the elastic scattering is comparable with the SOC scattering. For the weak SOC approximation, the influence of $B_e$ on $B_{so}$ is very weak and could be ignored, as the case of $x$ = 0.2 (Figure S4, Supporting Information). However, the failure of such approximation at $x$ = 0.225 and 0.25 implies a significant influence of $B_e$ on $B_{so}$. This could also explain the decrease of maximum spin splitting energy from 5.5 meV to 1.6 meV with the increasing Mn doping from 0.2 to 0.25. Moreover, for $x$ = 0.3 (Figure S5(a)), only classical positive MR is observed at different bias voltages before becoming insulator, indicating that the $B_e$ is lower than $B_i$ and the WAL/WL does not appear.

The gate-tunable transport properties of different electronic states at the diluted oxide interfaces with $0.2 \leq x \leq 0.3$ are summarized in Figure 5(b). Here, we observe the WAL/WL at $3.66 \times 10^{13}$ cm$^{-2}$ $\leq n_s \leq 0.88 \times 10^{13}$ cm$^{-2}$ in the low mobility region. With the increase of mobility, the anomalous magnetoresistance caused by WAL/WL changes to classical MR, implying the transition of transport from localized states to extended states.

## IV. CONCLUSIONS

In conclusion, we have systematically investigated the electrostatic modulation of the transport properties of LAMO/STO heterointerfaces. Regardless the carrier density, at the interfaces with lower mobility, where the $B_e$ is high enough and the weak SOC approximation is available, the influence of $B_e$ on $B_{so}$ is negligible. However, for higher mobility samples, where $B_e$ is comparable with the $B_{so}$, a significant suppression of spin splitting energy occurs. Our observations not only mapped the evolution of spin-orbit coupling as the function of carrier density, when the interfaces change from extended state to WAL, and then to WL, but also realized the gate-tunable spin-polarization of 2DELs. These results make the LAMO/STO system an intriguing platform for future oxide spintronics.

**Acknowledgements**

**References**

[1] A.D. Caviglia, M. Gabay, S. Gariglio, N. Reyren, C. Cancellieri, and J.M. Triscone, Phys. Rev. Lett. **104**, 126803 (2010).

[2] H. Nakamura, T. Koga, and T. Kimura, Phys. Rev. Lett. **108**, 206601 (2012).

[3] H. Liang, L. Cheng, L. Wei, Z. Luo, G. Yu, C. Zeng, and Z. Zhang, Phys. Rev. B - Condens. Matter Mater. Phys. **92**, 075309 (2015).

[4] W. Niu, Y. Zhang, Y. Gan, D. V. Christensen, M. V. Soosten, E.J. Garcia-Suarez, A. Riisager, X. Wang, Y. Xu, R. Zhang, N. Pryds, and Y. Chen, Nano Lett. **17**, 6878 (2017).

[5] A. Brinkman, M. Huijben, M. Van Zalk, J. Huijben, U. Zeitler, J.C. Maan, W.G. Van Der Wiel, G. Rijnders, D.H.A. Blank, and H. Hilgenkamp, Nat. Mater. **6**, 493 (2007).

[6] B. Kalisky, J.A. Bert, B.B. Klopfer, C. Bell, H.K. Sato, M. Hosoda, Y. Hikita, H.Y. Hwang, and K.A. Moler, Nat. Commun. **3**, 922 (2012).

[7] D. V. Christensen, Y. Frenkel, Y.Z. Chen, Y.W. Xie, Z.Y. Chen, Y. Hikita, A. Smith, L. Klein, H.Y. Hwang, N. Pryds, and B. Kalisky, Nat. Phys. **15**, 269 (2019).

[8] A.D. Caviglia, S. Gariglio, N. Reyren, D. Jaccard, T. Schneider, M. Gabay, S. Thiel, G. Hammerl, J. Mannhart, and J.M. Triscone, Nature **456**, 624 (2008).

[9] A. Joshua, S. Pecker, J. Ruhman, E. Altman, and S. Ilani, Nat. Commun. **3**, 1129 (2012).

[10] G. Herranz, G. Singh, N. Bergeal, A. Jouan, J. Lesueur, J. Gázquez, M. Varela, M. Scigaj, N. Dix, F. Sánchez, and J. Fontcuberta, Nat. Commun. **6**, 6028 (2015).

[11] F. Trier, G.E.D.K. Prawiroatmodjo, Z. Zhong, D.V. Christensen, M. Von Soosten, A. Bhowmik, J.M.G. Lastra, Y. Chen, T.S. Jespersen, and N. Pryds, Phys. Rev. Lett. **117**, 96804

(2016).

[12] C.H. Ahn, J.M. Triscone, and J. Mannhart, Nature **424**, 1015 (2003).

[13] D. V. Christensen, F. Trier, W. Niu, Y.L. Gan, Y. Zhang, T.S. Jespersen, Y.Z. Chen, and N. Pryds, Adv. Mater. Interfaces **Submitting**, (2019).

[14] F. Trier, D. V. Christensen, and N. Pryds, J. Phys. D. Appl. Phys. **51**, 293002 (2018).

[15] S. Thiel, G. Hammerl, A. Schmehl, C.W. Schneider, and J. Mannhart, Science. **313**, 1942 (2006).

[16] D. Stornaiuolo, C. Cantoni, G.M. De Luca, R. Di Capua, E. Di Gennaro, G. Ghiringhelli, B. Jouault, D. Marrè, D. Massarotti, F.M. Granozio, I. Pallecchi, C. Piamonteze, S. Rusponi, F. Tafuri, and M. Salluzzo, Nat. Mater. **15**, 278 (2016).

[17] J. Mannhart and D.G. Schlom, Science. **327**, 1607 (2010).

[18] J. Varignon, L. Vila, A. Barthélémy, and M. Bibes, Nat. Phys. **14**, 322 (2018).

[19] Y.Z. Chen, F. Trier, T. Wijnands, R.J. Green, N. Gauquelin, R. Egoavil, D. V. Christensen, G. Koster, M. Huijben, N. Bovet, S. Macke, F. He, R. Sutarto, N.H. Andersen, J.A. Sulpizio, M. Honig, G.E.D.K. Prawiroatmodjo, T.S. Jespersen, S. Linderoth, S. Ilani, J. Verbeeck, G. Van Tendeloo, G. Rijnders, G.A. Sawatzky, and N. Pryds, Nat. Mater. **14**, 801 (2015).

[20] T. Fix, F. Schoofs, J.L. MacManus-Driscoll, and M.G. Blamire, Phys. Rev. Lett. **103**, 166802 (2009).

[21] Y.L. Gan, D.V. Christensen, Y. Zhang, H.R. Zhang, D. Krishnan, Z.C. Zhong, W. Niu, D.J. Carrad, K. Norrman, M. von Soosten, T.S. Jespersen, B. Shen, N. Gauquelin, J. Verbeeck, J.R. Sun, N. Pryds, and Y.Z. Chen, Adv. Mater. **31**, 1805970 (2019).

[22] A.E.M. Smink, J.C. De Boer, M.P. Stehno, A. Brinkman, W.G. Van Der Wiel, and H. Hilgenkamp, Phys. Rev. Lett. **118**, 106401 (2017).

[23] G. Bergmann, Phys. Rev. B **28**, 2914 (1983).

[24] W. Liu, S. Gariglio, A. Fête, D. Li, M. Boselli, D. Stornaiuolo, and J.M. Triscone, APL Mater. **3**, 62805 (2015).

[25] S. Maekawa and H. Fukuyama, J. Phys. Soc. Japan **50**, 2516 (1981).

[26] F. Trier, G.E.D.K. Prawiroatmodjo, M. Von Soosten, D. V. Christensen, T.S. Jespersen, Y.Z. Chen, and N. Pryds, Appl. Phys. Lett. **107**, 191604 (2015).

[27] M. Gabay and J.M. Triscone, Nat. Phys. **9**, 610 (2013).

[28] Z. Chen, H. Yuan, Y. Xie, D. Lu, H. Inoue, Y. Hikita, C. Bell, and H.Y. Hwang, Nano Lett. **16**, 6130 (2016).

[29] F. Gunkel, C. Bell, H. Inoue, B. Kim, A.G. Swartz, T.A. Merz, Y. Hikita, S. Harashima, H.K. Sato, M. Minohara, S. Hoffmann-Eifert, R. Dittmann, and H.Y. Hwang, Phys. Rev. X **6**, 031035 (2016).

[30] C. Bell, S. Harashima, Y. Kozuka, M. Kim, B.G. Kim, Y. Hikita, and H.Y. Hwang, Phys. Rev. Lett. **103**, 226802 (2009).

[31] Y. Zhang, Y.L. Gan, D.V. Christensen, H. Zhang, A. Zakharova, C. Piamonteze, S.H. Wang, N. Pryds, B.G. Shen, J.R. Sun, and Y.Z. Chen, **unpulished**, (n.d.).

[32] M. Kohda, T. Bergsten, and J. Nitta, J. Phys. Soc. Japan **77**, 1 (2008).

[33] T. Jungwirth and J. Wunderlich, Nat. Nanotechnol. **9**, 662 (2014).

[34] G. Bergmann, Solid State Commun. **42**, 815 (1982).

[35] S. HIKAMI, A.I. LARKIN, and Y. NAGAOKA, Prog. Theor. Phys. **63**, 707 (1980).

[36] B.L. Al'tshuler, A.G. Aronov, A.I. Larkin, and D.E. Khmel'nitskii, Sov. Phys. JETP **54**, 0411 (1981).

[37] S. Iordanskii, Y. Lyanda-Geller, and G. Pikus, Sov. J. Exp. Theor. Phys. Lett. **60**, 206 (1994).

[38] A. Punnoose, Appl. Phys. Lett. **88**, 252113 (2006).

[39] F. Duan and J. GuojunF, *Introduction to Condensed Matter Physics* (World Scientific Publishing Co. Pte. Ltd., 2005).

[40] A. McCollam, S. Wenderich, M.K. Kruize, V.K. Guduru, H.J.A. Molegraaf, M. Huijben, G. Koster, D.H.A. Blank, G. Rijnders, A. Brinkman, H. Hilgenkamp, U. Zeitler, and J.C. Maan, APL Mater. **2**, 22102 (2014).

[41] Z. Zhong, A. Tóth, and K. Held, Phys. Rev. B - Condens. Matter Mater. Phys. **87**, 161102(R) (2013).

**Table**

Table 1. Transport properties for different as-grown states of LaAl$_{1-x}$Mn$_x$O$_3$/SrTiO$_3$ interfaces

| Mn component ($x$) | $n_{tot}$ (cm$^{-2}$) | $\mu_{tot}$ (cm$^2$/Vs) | Conduction band | Magnetism |
|---|---|---|---|---|
| 0.2 | 3.90×10$^{13}$ | 568 | $d_{xy} + d_{xz}/d_{yz}$ | × |
| 0.225 | 2.54×10$^{13}$ | 597 | $d_{xy}$ | × |
| 0.25 | 2.14×10$^{13}$ | 1566 | $d_{xy}$ | × |
| 0.3 | 0.74×10$^{13}$ | 1613 | $d_{xy}$ | √ |

**Figures**

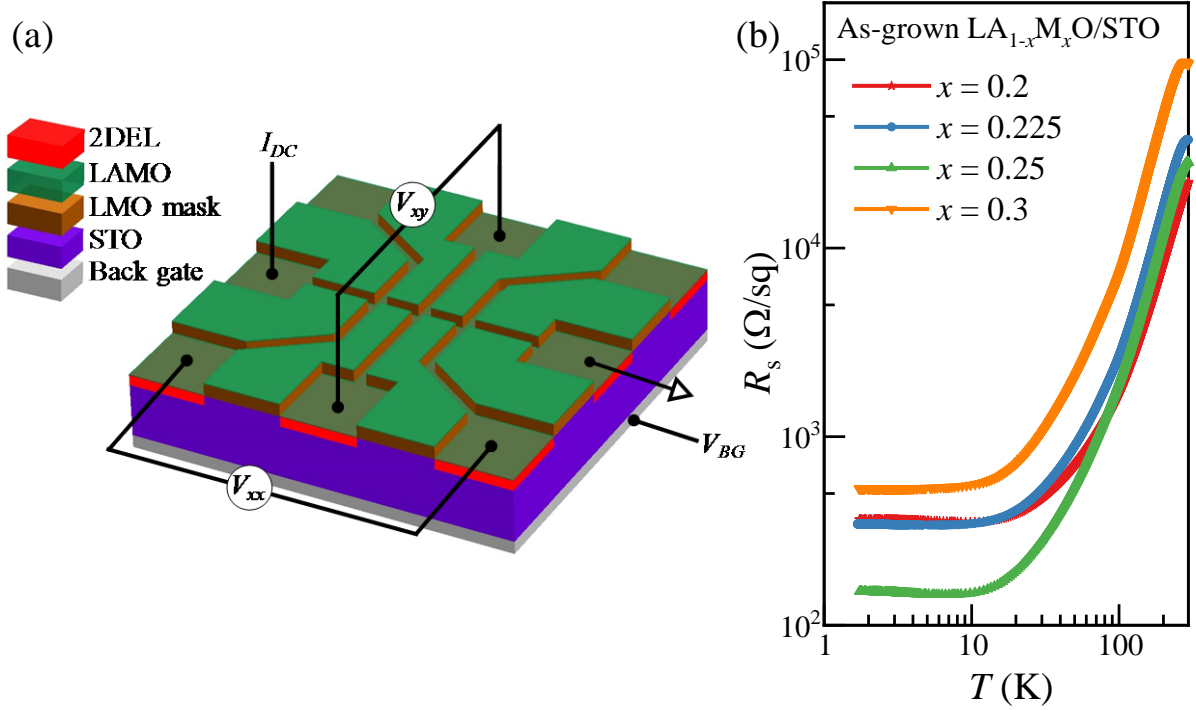

(a)

2DEL
LAMO
LMO mask
STO
Back gate

$I_{DC}$

$V_{xy}$

$V_{xx}$

$V_{BG}$

(b) As-grown LA$_{1-x}$M$_x$O/STO

$x = 0.2$
$x = 0.225$
$x = 0.25$
$x = 0.3$

$R_s$ ($\Omega$/sq)

$T$ (K)

**FIG. 1.** (a) Sketch of the LaAl$_{1-x}$Mn$_x$O$_3$/SrTiO$_3$ devices under the various back-gate voltages. (b) Temperature dependence of the sheet resistance ($R_s$) for as-grown states with different Mn doping concentrations ($x$)

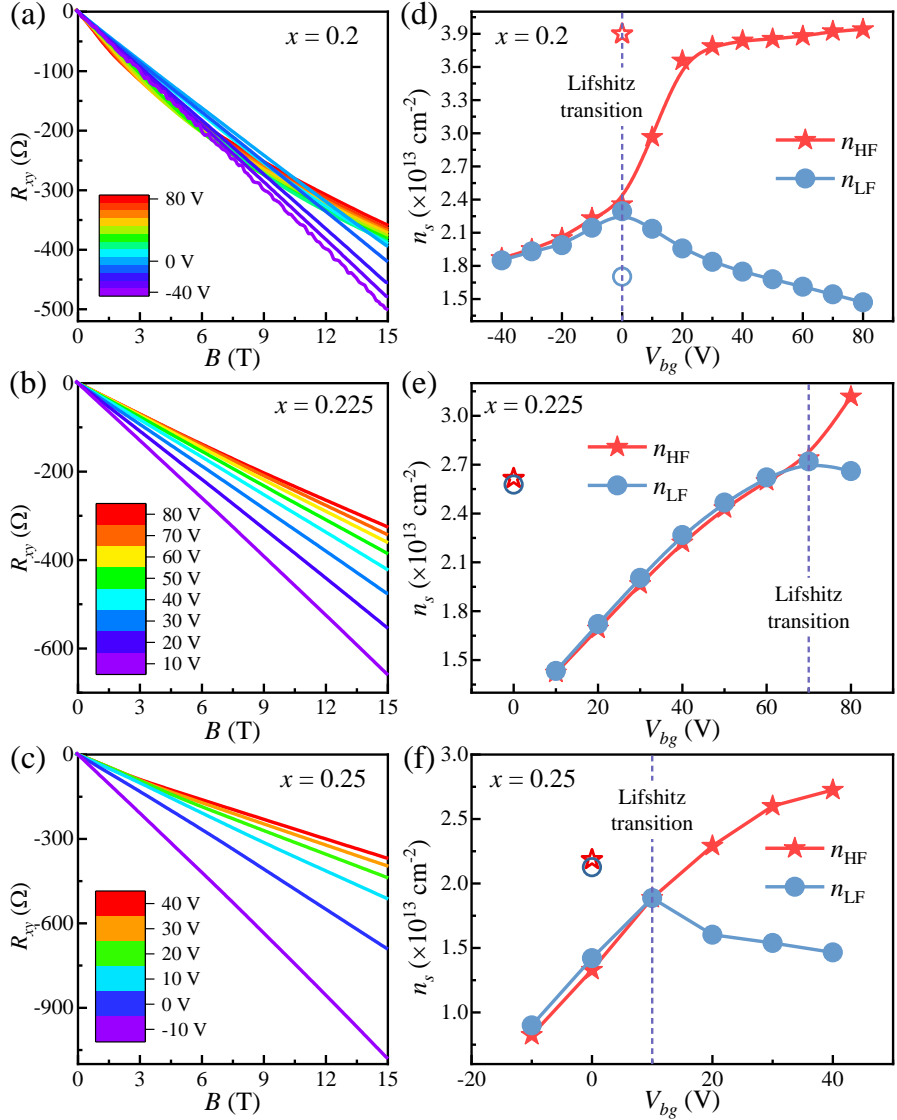

**FIG. 2.** Hall resistance and carrier density as a function of the gate voltage at interfaces with $x$ = 0.2, 0.225 and 0.25. (a), (b) and (c) show the $B$-dependent Hall resistance for various gate voltages for LaAl$_{1-x}$Mn$_x$O$_3$/SrTiO$_3$ heterointerfaces with $x$ = 0.2, 0.225 and 0.25 respectively. (d), (e) and (f) display the $n_{LF}$ and $n_{HF}$ extracted from low field (0 T) and high field (15 T) of $R_{xy}$-$B$ curves as a function of gate voltage for different Mn-doping level. The open scatter plots show the as-grown carrier density for each states.

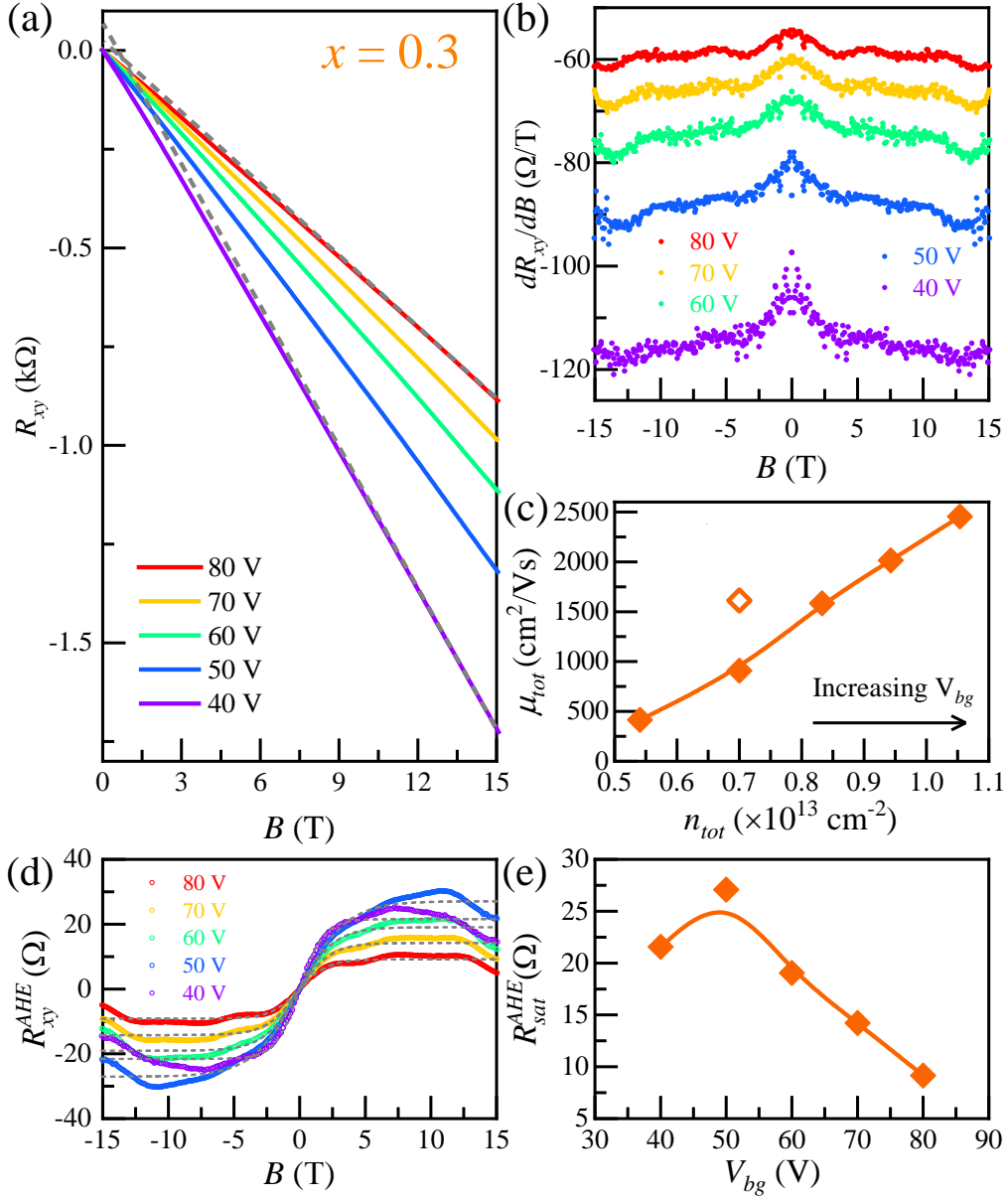

**FIG. 3.** Gate-tunable anomalous Hall effect (AHE), carrier density and mobility at spin-polarized interface with $x = 0.3$. (a) The nonlinear $R_{xy}$ versus $B$ from 40 V to 80 V. (b) The corresponding evolution of Hall coefficient $R_H = \mathrm{d}R_{xy}/\mathrm{d}B$. (c) The carrier density and mobility extracted from the fitting ordinary Hall resistance. The open rhombus shows the as-grown carrier density and mobility for $x = 0.3$. (d) Magnetic field dependent $R_{xy}^{AHE}$ under different $V_{bg}$. The dash line is the fitting data of anomalous Hall resistance. (e) The saturated Hall resistance $R_{sat}^{AHE}$ as a function of $V_{bg}$.

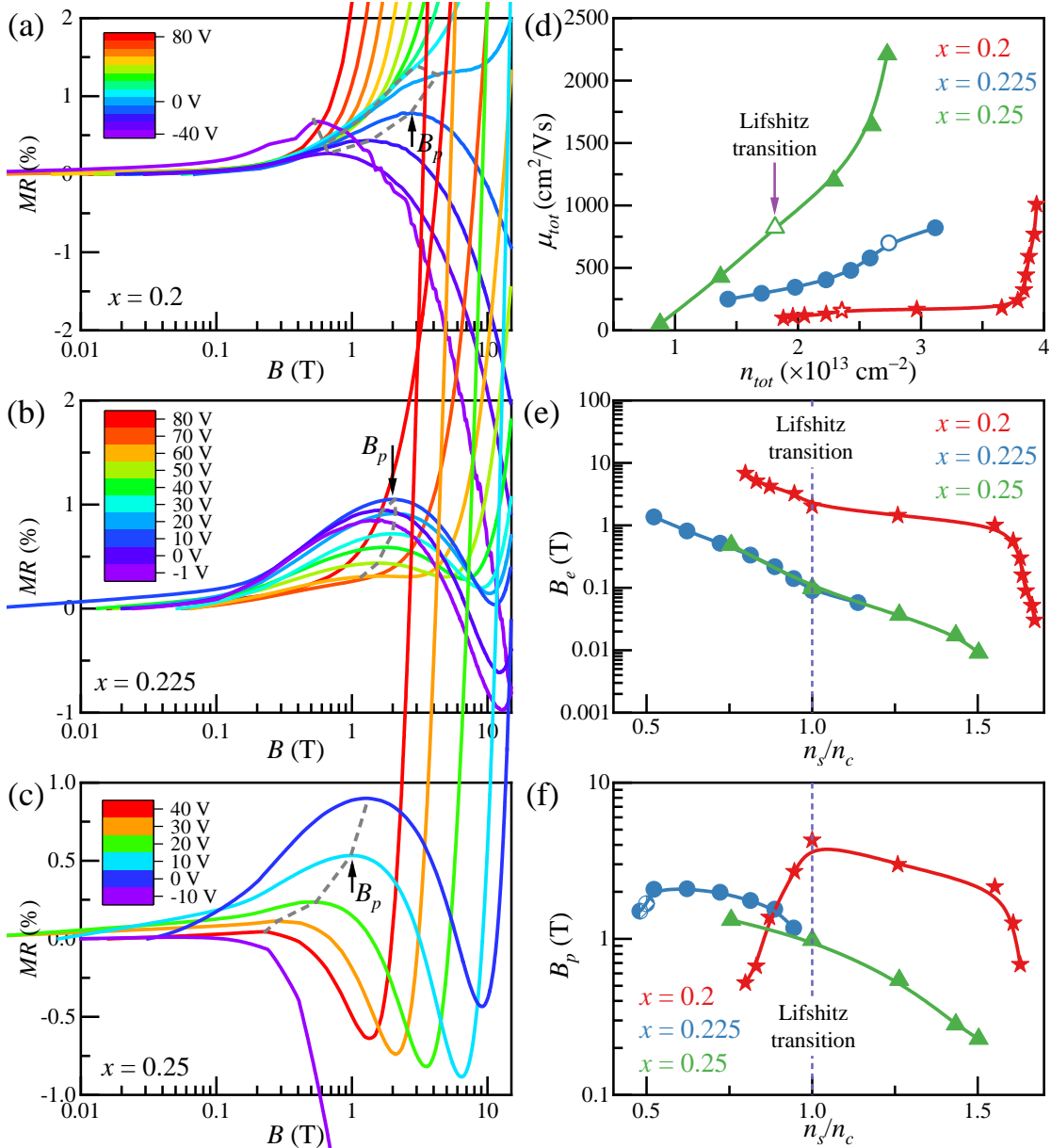

**FIG. 4.** Analysis of the magnetoresistance data at interfaces with $x = 0.2$, 0.225 and 0.25. (a), (b) and (c) show the field-dependent magnetoresistance measurements carried out under different back-gate voltages for $x = 0.2$, 0.225 and 0.25, respectively. The grey dash lines mark the position of $B_p$. In order to highlight the anomalous magnetoresistance at the low magnetic field, the MR curves are shown in a semilogarithmic plot of $B$. (d) The evolutions of mobility as a function of carrier density. The open star, circle and triangle are the corresponding Lifshitz transitions for $x = 0.2$, 0.225 and 0.25, respectively. (e) The evolutions of elastic scattering fields reduced by $n_{tot}$ and $\mu_{tot}$ as the function of carrier density normalized by the Lifshitz density ($n_c$). (f) The evolutions of $B_p$ extracted from (a-c) as the function of carrier density normalized by the Lifshitz density ($n_c$). The left and right half-filled circles mark the interfaces with $x = 0.225$ at -1 V and 0 V, where the Hall resistance is undetectable and their normalized carrier density is lower than 0.5.

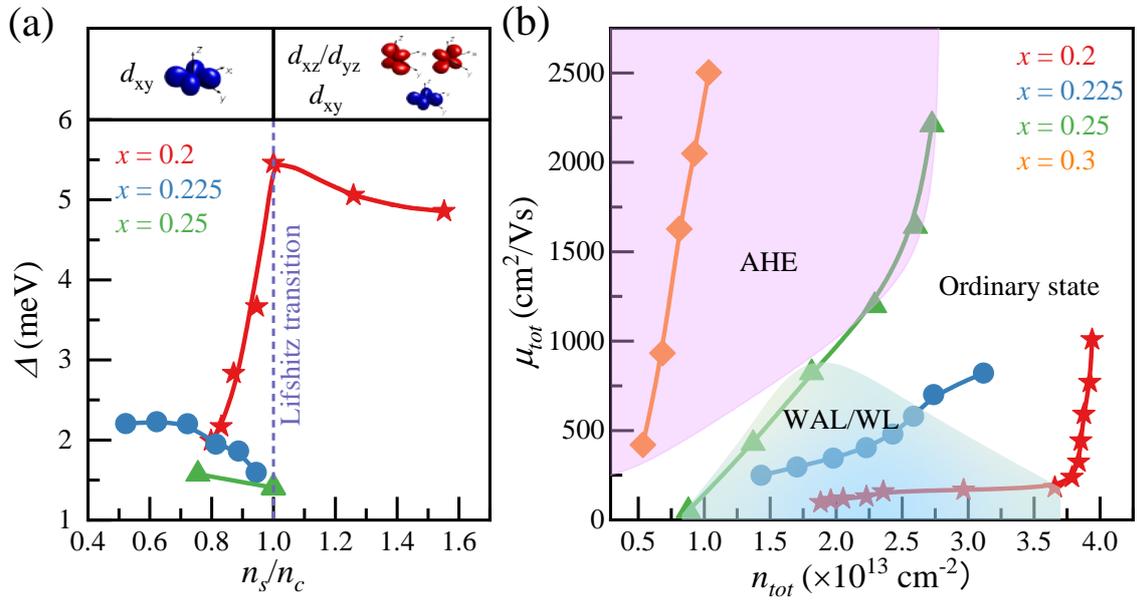

**FIG. 5.** (a) The spin splitting energy as functions of the sheet carrier density ($n_s$) normalized by the Lifshitz density ($n_c$) for the LaAl$_{1-x}$Mn$_x$O$_3$/SrTiO$_3$ interfaces with $x = 0.2$, 0.225 and 0.25. (b) Carrier mobility as a function of carrier density for the four samples. The regions for anomalous Hall effect and robust weak (anti)localization are summarized.